\documentclass[aps,twocolumn,showpacs]{revtex4}
\usepackage{graphicx}

\begin{document}
\title{Optimal copying of entangled two-qubit states}
\author{J. Novotn\'y$^{(1)}$, G. Alber$^{(2)}$, I. Jex$^{(1)}$}
\affiliation{$^{(1)}$~Department of Physics, FJFI \v CVUT,
B\v rehov\'a 7,
115 19 Praha 1 - Star\'e M\v{e}sto, Czech Republic\\
$^{(2)}$~Institut f\"ur Angewandte Physik, Technische Universit\"at Darmstadt, D-64289 Darmstadt, Germany}
\date{today}
\begin{abstract}
We investigate
the problem of copying pure two-qubit states of a given degree of entanglement
in an optimal way. Completely positive covariant quantum operations are constructed which
maximize the fidelity of the output states with respect to two
separable copies. These optimal copying processes hint at the
intricate relationship between fundamental laws of quantum theory and entanglement.
\end{abstract}
\maketitle

\section{Introduction}


The optimal copying (cloning) of quantum states is an elementary process of central interest in
quantum information processing \cite{general}.
As arbitrary quantum states cannot be copied
perfectly, the interesting question arises as to which extent a quantum process can perform this task
in an optimal way.
Recently, much work has been devoted to the copying of pure quantum states \cite{copying1,Gisin,Niu,Werner,Cerf1,Cerf2,Braunstein,Cerf3,Cerfetal}.
These investigations
demonstrate that the maximal fidelity which can be achieved in an optimal copying process
depends on characteristic properties of the set of states which are to be copied.

Motivated by the important role entanglement is playing in the
area of quantum information processing in this paper we address
the question of copying pure entangled two-qubit states in an
optimal way. Though much work has already been devoted to the
copying  of pure single particle states, first investigations
addressing the problem of copying entanglement have been performed
only recently \cite{Cerfetal}. In this latter work it was
demonstrated that entanglement cannot be copied perfectly. Thus,
if one can find a quantum operation which perfectly duplicates
entanglement of all maximally entangled qubit pairs, it
necessarily cannot respect separability of the two identical
copies produced.
Furthermore, for the special
case of maximally entangled two-qubit states first copying processes
were constructed which maximize the fidelity of each two-qubit copy separately.

In this paper we address the general problem of copying pure two-qubit states of an arbitrarily given degree of entanglement
in an optimal way.
In particular, we are interested in constructing completely positive quantum operations which
do not only copy the pure entangled input state but which also guarantee separability of the resulting
two-qubit copies in an optimal way.
Our motivation for restricting our investigation to two-qubit
input states is twofold. Firstly, qubit states still play a
dominant role in the area of quantum information processing.
Secondly, it is expected that in this simplest case the intricate
relationship between entanglement and limits imposed on quantum
copying processes by the fundamental laws of quantum theory are
exposed in a particularly transparent way.

This paper is organized as follows:
In Sec. II
we briefly recapitulate the basic relation between optimal copying processes
and corresponding covariant quantum processes
which
maximize the fidelity of the output states with respect to separable copies.
In Sec. III the most general covariant quantum processes are constructed which are consistent with the linear character of quantum maps
and which copy  arbitrary pure two-qubit quantum states of a given degree of entanglement.
In Sec. IV the additional constraints are investigated which result from the
positivity of these quantum maps.  Based on these results the parameters of the optimal copying processes are determined.
In Sec. V it is shown that these optimal covariant quantum maps can be realized by completely positive deterministic quantum operations.
Thus, they may be implemented
by an appropriate unitary transformation and a subsequent measurement process involving
additional ancilla qubits.
Basic physical properties of the resulting optimally copied states are discussed in Sec.VI.

\section{Optimal copying of entangled two-qubit states}

In this section basic connections between optimal quantum
mechanical copying processes and covariant quantum processes are
summarized and specialized to the problem of copying pure
bipartite entangled two-qubit states in an optimal way.

In order to put the problem into perspective let us consider two
distinguishable spin-$1/2$ particles (qubits). Their associated
four dimensional Hilbert space $\mathcal{H}$ can be decomposed
into classes  of pure two-qubit states $\Omega_{\alpha}$ of a
given degree of entanglement $\alpha$ (the parameter $\alpha$
determines the amount of entanglement in the bipartite system
described by states from $\Omega_\alpha$). These classes are
represented by the sets
\begin{widetext}
\begin{equation}
\Omega_{\alpha}= \Big\{ \big(U_{1} \otimes U_{2}\big) \big(
\alpha |\uparrow\rangle_1 \otimes  |\uparrow \ \rangle_2 + \sqrt{1-\alpha^2} |\downarrow\rangle_1 \otimes |\downarrow \rangle_2
\big) \Big| U_{1}, U_{2} \in {SU}(2)  \Big\}.
\label{Omega}
\end{equation}
\end{widetext}
Thereby the parameter $\alpha$ ($0 \leq \alpha \leq 1$)
characterizes the degree of entanglement of the pure states in a given class $\Omega_{\alpha}$
and the kets
$|\uparrow \rangle$ and $|\downarrow \rangle $ constitute an
orthonormal basis of the
two-dimensional single-qubit Hilbert spaces of each of the qubits (distinguished by the subscripts $1$ and $2$).
Relation (\ref{Omega}) takes into account that
local unitary operations of the form $U_1\otimes U_2$ are the most
general transformations which leave the
degree (measure) of entanglement $\alpha$ of a bipartite quantum state invariant.
Due to the symmetry relation
$\Omega_{\alpha} = \Omega_{ \sqrt{1- \alpha^2}}$
we can restrict our subsequent discussion to the parameter range
$0\leq \alpha \leq 1/\sqrt{2}$. Note that in the special case $\alpha =0$ the two-qubit state
is separable whereas in the opposite extreme case $\alpha = 1/\sqrt{2}$ it is maximally entangled.
Furthermore, it should be noted that each class $\Omega_\alpha$ contains an
orthonormal basis of $\mathcal{H}$.

We are interested in constructing quantum processes $T_{\alpha}$ which copy an arbitrary pure two-qubit
state, say
$|\psi\rangle \in \Omega_{\alpha}$,
in an optimal way, i.e.
\begin{equation}
T_{\alpha} : \rho_0 \equiv \rho_{in} \otimes \rho_{{ref}} \longrightarrow
\rho_{out},
\label{map}
\end{equation}
with $\rho_{in} = |\psi\rangle \langle \psi |$ denoting the
density operator of the the input state. The resulting four-qubit
output state is denoted by $\rho_{out}$. The appropriately chosen
two-qubit quantum state $\rho_{{ref}}$ characterizes the initial
state of the copying device which is independent of the input
state. According to the fundamental laws of quantum theory the
quantum map $T_{\alpha}$  has to be linear and completely positive
\cite{Kraus,Peres,Nielsen}.

The fidelity
$F = \langle \psi | \otimes \langle \psi  |\rho_{out} | \psi
\rangle \otimes | \psi \rangle $
constitutes
a convenient quantitative measure of how close an output state
$\rho_{out}$ resembles two ideal separable copies of the original input state $\rho_{in} = |\psi\rangle \langle \psi |$.
Consequently, the
smallest achievable
fidelity, i.e.
\begin{equation}
{\mathcal{L}}(T_{\alpha}) = \inf_{| \psi \rangle \in
\Omega_{\alpha}} \langle \psi | \otimes \langle \psi | \rho_{out}
| \psi \rangle \otimes | \psi \rangle,
\label{fid}
\end{equation}
characterizes the quality of a copying process for a given class
of entangled input states $ | \psi\rangle \in \Omega_{\alpha}$.
Thus, constructing an optimal copying process is equivalent to
maximizing ${\mathcal{L}}(T_{\alpha})$ over all possible quantum
processes. Let us denote this optimal fidelity by
$\widehat{{\mathcal{L}}_{\alpha}} \equiv
\sup_{T_{\alpha}}{\mathcal{L}}(T_{\alpha})$. It has been shown
\cite{Gisin,Werner} that for any optimal quantum copying process
$\widehat{T_{\alpha}}$ one can always find an equivalent covariant
quantum process
with the
characteristic property
\begin{eqnarray}
\label{covariant} \rho_{\rm out}[U_1\otimes U_2 \rho_{\rm
in}U^{\dagger}_1\otimes U^{\dagger}_2] &=&
{\cal U}
\rho_{\rm out}( \rho_{in} )
{\cal U}^{\dagger}
\label{covariance}
\end{eqnarray}
with ${\cal U} =
U_1\otimes U_2\otimes U_1\otimes U_2$.
Thus, this equivalent covariant quantum process
yields the same optimal fidelity $\widehat{{\mathcal{L}}_{\alpha}}$ for all possible
two-qubit input states $|\psi\rangle \in \Omega_{\alpha}$.
Thereby, $U_1, U_2 \in {SU}_2$  are arbitrary unitary one-qubit transformations.
A proof of this theorem is sketched in Appendix A.
This observation allows us to restrict our further search for optimal copying processes of entangled pure two-qubit states
to covariant quantum processes which maximize the fidelity of Eq.(\ref{fid}).

At this point we want to emphasize that in this work we are concerned with optimal entanglement processes which maximize
the fidelity of Eq.(\ref{fid}). Thus, the covariant copying processes we are looking for are designed for producing
output states of the form $| \psi \rangle \otimes | \psi \rangle$, i.e. two separable pairs of pure entangled two-qubit states,
with the highest possible probability for all possible two-qubit input states
$| \psi \rangle \in \Omega_{\alpha}$. As we are focusing on separable copies of the input states
these processes do not necessarily also maximize
the fidelity $F'$ of the output state with respect to each copy separately, i.e. with respect to
$F'= Tr\{| \psi \rangle \langle \psi| \otimes {\bf 1} \rho_{out}(| \psi \rangle \langle \psi|)\}$. These latter processes
were
studied in Ref.\cite{Cerfetal}, for example, for the special case of maximally entangled pure input states, i.e. $\alpha = 1/\sqrt{2}$.

\section{Covariant linear quantum processes}

In this section all possible covariant copying processes are
constructed which are consistent with the linear character of
general quantum maps of the form of Eq.(\ref{map}).

In view of the covariance condition (\ref{covariance}) all
possible quantum maps of the form (\ref{map}) can be characterized
by the output states $\rho_{out}(\rho_{in})$ which originate from
one arbitrarily chosen pure input state, say $|\psi\rangle
=\alpha|\uparrow\rangle_1\otimes|\uparrow\rangle_2 +
\sqrt{1-\alpha^2}|\downarrow\rangle_1\otimes|\downarrow\rangle_2$
with $0\leq \alpha \leq 1/\sqrt{2}$. In order to fulfill
Eq.(\ref{covariance}) the two-qubit reference state $\rho_{ref}$
of Eq.(\ref{map}) has to be invariant under arbitrary local
unitary transformations of the form $U_1\otimes U_2$. Therefore,
the initial state of the covariant quantum map is of the form
\cite{Werner}
\begin{eqnarray}
\rho_{0}= \rho_{in} \otimes \frac{1}{4}{\bf 1}\equiv |\psi\rangle
\langle \psi |\otimes \frac{1}{4}{\bf 1} . \label{inputdm}
\end{eqnarray}

In order to implement the covariance condition of Eq.(\ref{covariance}) it is convenient
to decompose this quantum state
into irreducible two-qubit tensor operators $T^{(1,3)}(J',J)_{KQ}$ and $T^{(2,4)}(J',J)_{KQ}$
\cite{Blum,Biedenharn}
with respect to qubits one and three on the one hand
and qubits two and four on the other hand.
Performing an
arbitrary unitary transformation of the form
$U_1 \otimes U_2 \otimes U_1 \otimes U_2$ with $U_1, U_2 \in SU_{2}$, for example,
a product of such tensor operators transforms according to
\begin{widetext}
\begin{eqnarray}
\label{irred}
{\cal U}
~T^{(1,3)}(J_1'J_1)_{K_1Q_1} \otimes T^{(2,4)}(J_2'J_2)_{K_2Q_2}
{\cal U}^{\dagger}
&=&
\sum_{q_1,q_2}
D(U_1)_{q_1Q_1}^{(K_1)}
D(U_2)_{q_2Q_2}^{(K_2)}
T^{(1,3)}(J_1'J_1)_{K_1q_1}\otimes
T^{(2,4)}(J_2'J_2)_{K_2q_2}
\nonumber
\end{eqnarray}
\end{widetext}
with ${\cal U} = U_1 \otimes U_2 \otimes U_1 \otimes U_2 $.
Thereby, $D(U_j)$ ($j=1,2$) denote the relevant rotation operators
and $D(U_j)_{q_jQ_j}^{(K_j)}$ are their associated rotation
matrices \cite{Biedenharn}. The quantum numbers $J_j, J'_j$ denote
the total angular momenta of the relevant two-qubit quantum states
and the parameters $K_j$ indicate the irreducible subspaces of the
relevant representations. For the sake of convenience some basic
relations of these irreducible two-qubit tensor operators are
summarized in Appendix B. It is apparent from Eq.(\ref{irred})
that an arbitrary unitary transformation of the form $U_1 \otimes
U_2 \otimes U_1 \otimes U_2$ with $U_1, U_2 \in SU_{2}$ mixes the
parameters $q_1$ and $q_2$ within each irreducible representation
separately. In terms of these irreducible tensor operators an
arbitrary initial state $\rho_0$ of the form of Eq.(\ref{inputdm})
can be decomposed according to (compare with Eq.(\ref{decompose}))
\begin{widetext}
\begin{eqnarray}
\rho_{0} &=&
\sum_{j_1,...,j_4,K,Q,K',Q'}
T^{(1,3)}(j_1,j_3)_{KQ}\otimes T^{(2,4)}(j_2,j_4)_{K'Q'}
\langle T^{(1,3)\dagger}(j_1,j_3)_{KQ}T^{(2,4)\dagger}(j_2,j_4)_{K'Q'} \rangle
\label{output}
\end{eqnarray}
\end{widetext}
with
the expansion coefficients
\begin{widetext}
\begin{eqnarray}
&&
\langle T^{(1,3)\dagger}(j_1,j_3)_{KQ}T^{(2,4)\dagger}(j_2,j_4)_{K'Q'}\rangle \equiv
Tr \big\{ (T^{(1,3)\dagger}(j_1,j_3)_{KQ}\otimes
T^{(2,4)\dagger}(j_2,j_4)_{K'Q'}) \rho_{0} \big\}.
\end{eqnarray}
\end{widetext}
Thereby, $Tr$ denotes the trace over the four-qubit Hilbert space of the system- and device qubits.
In view of the basic
transformation property of Eq.(\ref{irred})
the most general output state resulting from a linear and covariant quantum map
is given by
\begin{widetext}
\begin{eqnarray}
\rho_{out}(\rho_{in}) &=& \sum_{j_1,...,j_4,K,Q,K',Q'}
\alpha(j_1,j_3,j_2,j_4)_{KK'}
~T^{(1,3)}(j_1,j_3)_{KQ}T^{(2,4)}(j_2,j_4)_{K'Q'}
 \langle T^{(1,3)\dagger}(j_1,j_3)_{KQ}T^{(2,4)\dagger}(j_2,j_4)_{K'Q'}
\rangle. \nonumber\\
&&\label{output1}
\end{eqnarray}
\end{widetext}
According to the fundamental laws of quantum theory the unknown
coefficients $\alpha(j_1,j_3,j_2,j_4)_{KK'}$ are necessarily
restricted by the fact that $\rho_{out}$ has to be a non-negative
operator. In particular, being a Hermitian operator the output
state $\rho_{out}$ has to fulfill the relations
\begin{equation}
\alpha(j_1,j_3,j_2,j_4)_{KK^{'}}=\alpha(j_3,j_1,j_4,j_2)^*_{KK^{'}}.
\label{hermit}
\end{equation}
Further restrictions on these unknown coefficients are obtained from the explicit form of the
input state $\rho_0$, i.e.
\begin{widetext}
\begin{eqnarray}
\rho_{0}&=&
 \frac{|\alpha|^{2}}{4} \Big\{
\frac{1}{\sqrt{2}} T^{(1,3)}(1,1)_{10} + \frac{3}{2}T^{(1,3)}(1,1)_{00} +
\frac{1}{2}T^{(1,3)}(0,0)_{00} - \frac{1}{2}T^{(1,3)}(0,1)_{1,0} + \frac{1}{2}
T^{(1,3)}(1,0)_{10} \Big\} \otimes \nonumber \\
&&\hspace*{0.7cm}\Big\{ \frac{1}{\sqrt{2}} T^{(2,4)}(1,1)_{10} +
\frac{3}{2}T^{(2,4)}(1,1)_{00} +
\frac{1}{2}T^{(2,4)}(0,0)_{00} -
\frac{1}{2}T^{(2,4)}(0,1)_{1,0} + \frac{1}{2} T^{(2,4)}(1,0)_{10} \Big\}+ \nonumber\\&&
\frac{|\beta|^{2}}{4} \Big\{ \frac{-1}{\sqrt{2}} T^{(1,3)}(1,1)_{10} +
\frac{3}{2}T^{(1,3)}(1,1)_{00} +
\frac{1}{2}T^{(1,3)}(0,0)_{00} +
\frac{1}{2}T^{(1,3)}(0,1)_{1,0} - \frac{1}{2}
T^{(1,3)}(1,0)_{10} \Big\} \otimes \nonumber\\
&&\hspace*{0.7cm} \Big\{ \frac{-1}{\sqrt{2}} T^{(2,4)}(1,1)_{10} +
\frac{3}{2}T^{(2,4)}(1,1)_{00} +
\frac{1}{2}T^{(2,4)}(0,0)_{00} +
\frac{1}{2}T^{(2,4)}(0,1)_{1,0} - \frac{1}{2} T^{(2,4)}(1,0)_{10} \Big\} +\nonumber\\
&& \frac{\alpha \beta^{*}}{8} \Big\{ -\sqrt{2}T^{(1,3)}(1,1)_{11} +
T^{(1,3)}(0,1)_{11} -
T^{(1,3)}(1,0)_{11} \Big\} \otimes \nonumber\\
&&\hspace*{0.7cm}\Big\{
-\sqrt{2}T^{(2,4)}(1,1)_{11} + T^{(2,4)}(0,1)_{11} -
T^{(2,4)}(1,0)_{11} \Big\}+\nonumber\\&&
\frac{\alpha^{*} \beta}{8} \Big\{ \sqrt{2}T^{(1,3)}(1,1)_{1-1} -
T^{(1,3)}(0,1)_{1-1} +
T^{(1,3)}(1,0)_{1-1} \Big\}\otimes\nonumber\\
&&\hspace*{0.7cm}\Big\{ \sqrt{2}T^{(2,4)}(1,1)_{1-1} -
T^{(2,4)}(0,1)_{1-1} + T^{(2,4)}(1,0)_{1-1} \Big\}
\label{input}
\end{eqnarray}
\end{widetext}
with $\beta = \sqrt{1-\alpha^2}$.
Thus, according to Eq.(\ref{input}) the most general output state of Eq.(\ref{output1})
generally depends on
17 coefficients, namely
\begin{eqnarray}
&&
\alpha(1,1,1,1)_{11}=A_1, \alpha(1,1,1,1)_{10}=A_2,\nonumber\\&&
\alpha(1,1,1,0)_{11}=A_3, \alpha(1,1,0,0)_{10}=A_4,\nonumber\\&&
\alpha(1,1,1,1)_{01}=A_5,  \alpha(1,1,1,1)_{00}=A_6,\nonumber\\&&
\alpha(1,1,1,0)_{11}=A_7, \alpha(1,1,0,0)_{00}=A_8,\nonumber\\&&
\alpha(1,0,1,1)_{11}=A_9, \alpha(1,0,1,1)_{10}=A_{10},\nonumber\\&&
 \alpha(1,0,1,0)_{11}=A_{11}, \alpha(1,0,0,0)_{10}=A_{12},\nonumber\\&&
\alpha(0,0,1,1)_{01}=A_{13}, \alpha(0,0,1,1)_{00}=A_{14},\nonumber\\&&
 \alpha(0,0,1,0)_{01}=A_{15},\alpha(0,0,0,0)_{00}=A_{16},\nonumber\\&&
 \alpha(1,0,0,1)_{11}=A_{17}.
\label{parameters}
\end{eqnarray}
These parameters determine all linear covariant quantum processes
with a Hermitian output state $\rho_{out}(\rho_{in})$ provided the
coefficients $A_1, A_2, A_4, A_5, A_6, A_8, A_{13}, A_{14},
A_{16}$ are real-valued. The explicit form of the output state
$\rho_{out}(\rho_{in})$ is given in Appendix C (compare with
Eq.(\ref{matrix})). Proper normalization of the output state
requires $Tr\{\rho_{out}(\rho_{in})\} = 1$ which implies
\begin{equation}
\label{trace}
 \frac{1}{16} (9A_6 + 3A_8 + 3A_{14} + A_{16}) = 1.
\end{equation}

\section{Optimal covariant copying processes}

In this section
the special covariant
quantum processes are determined which copy pure entangled two-qubit states of a given degree of entanglement $\alpha$
with the highest possible fidelity.

For this purpose we start from
the most general output state
which is consistent
with the linear and covariant character of the copying process
as determined by
Eqs.(\ref{output1}), (\ref{hermit}), (\ref{trace}) and by an arbitrary combination of
the possible non-zero parameters $A_1,...,A_{17}$ of (\ref{parameters}).
The non-negativity of this output state on the one hand and the optimization of its fidelity
on the other hand impose further restrictions on these parameters.

The non-negativity of the output state implies that the inequality $\langle \chi |\rho_{out}(\rho_{in}) |\chi\rangle \geq 0$ has
to be fulfilled for  arbitrary pure four-qubit states $|\chi\rangle$.
As outlined in appendix C this condition gives rise to the set of inequalities
\begin{eqnarray}
&&
 A_6 \geq  0,  A_8 \geq 0,
 A_{14} \geq 0,
 A_{16} \geq 0,
|A_1| \leq A_6,\nonumber\\&&
\big| (2\alpha^2 -1)A_4 \big| \leq A_8,  \big| (2\alpha^2 -1)A_{13} \big| \leq A_{14},\nonumber \\&&
 \big| (2\alpha^2 -1)A_2 \big| \leq A_6, \big| (2\alpha^2 -1)A_5 \big| \leq A_6
\label{pos1}
\end{eqnarray}
and
\begin{eqnarray}
|A_{11}|^{2} \leq A_{16} A_{6},~
|A_{17}|^{2} &\leq& A_{14} A_{8},\nonumber\\
A_6 \mid (2\alpha^2 -1)(A_2 + A_5)\mid^2 &\leq&
(A_1 + A_6)^2A_6 -\label{pos2}\\
&& 8\alpha^2 (1-\alpha^2) A_1^2 (A_1 + A_6).\nonumber
\end{eqnarray}
In particular,
in the special case $A_1 = A_6\neq 0$  the last inequality of (\ref{pos2}) implies
\begin{eqnarray}
\mid (2\alpha^2 -1)(A_2 + A_5)\mid \leq
\sqrt{4 A_6^2 - 16 \alpha^2 (1-\alpha^2) A_6^2}.
\label{pos3}
\end{eqnarray}

The fidelity $F$ of the output state
$\rho_{out}(\rho_{in})$ of Eq.(\ref{output1})  with respect to the ideal pure two-qubit output state
$|\psi\rangle \otimes  |\psi\rangle$ with
$|\psi\rangle = \alpha|\uparrow\rangle_1\otimes|\uparrow\rangle_2  + \sqrt{1-\alpha^2} |\downarrow\rangle_1\otimes|\downarrow\rangle_2$
is given by
\begin{widetext}
\begin{eqnarray}
F&\equiv& \langle
\psi| \otimes \langle \psi| \rho_{out} |\psi \rangle \otimes |\psi
\rangle=
 \frac{1}{16} \Big\{ A_1 (1+2\alpha^2(1-\alpha^2)) +
(2\alpha^2
-1)^2(A_2 + A_5) +
A_6(1-\alpha^2(1-\alpha^2)) +
 \alpha^2 (1-\alpha^2) A_{16}
+\nonumber\\&&
 6\alpha^2(1- \alpha^2)  {\rm Re} A_{11} \Big\}.
\label{optimalfid}
\end{eqnarray}
\end{widetext}
Besides the parameter $\alpha$ determining the degree of
entanglement of the input state $|\psi\rangle$ this fidelity
depends on the six parameters $A_1, A_2, A_5, A_6, A_{11},
A_{16}$. An upper bound of this fidelity can be derived with the
help of the inequalities (\ref{pos2}), (\ref{pos3}) and with the
relation $A_{16} \leq 16 -9 A_6$ which is obtained from the
normalization condition (\ref{trace}), i.e.
\begin{widetext}
\begin{eqnarray}
\label{omezeni_fidelity}
 F &\leq & \frac{1}{16} \Big\{ A_6 (4-
16\alpha^2(1-\alpha^2)) + 16 \alpha^2(1-\alpha^2) +
6\alpha^2(1-\alpha^2)
\sqrt{A_6 (16 - 9A_6)}\Big\}.
\end{eqnarray}
\end{widetext}
This upper bound is
attained provided the conditions
$A_{16} = 16-9A_6$,
$A_{11} =\sqrt{A_6(16-9A_6)}$ and $A_1 = A_6 = (A_2+A_5)/2$
are fulfilled.
Maximizing the right hand side of Eq.(\ref{omezeni_fidelity})
with respect to the single parameter $A_6$ we finally arrive at the inequality
\begin{eqnarray}
\label{maximum} F\leq F_{max} &\equiv& \frac{2}{9} (1-
4\alpha^2(1-\alpha^2))(1+\sqrt{v}) +\nonumber\\&& \alpha^2(1-\alpha^2) (1+ \sqrt{1-v})
\label{upper}
\end{eqnarray}
with
\begin{eqnarray}
v&=&1-\frac{81\alpha^4(1-\alpha^2)^2}{145\alpha^4(1-\alpha^2)^2 - 32\alpha^2(1-\alpha^2)+4}.
\end{eqnarray}
This upper bound of relation (\ref{upper}) is reached
provided the parameters of the covariant copying process
fulfill the relations
\begin{eqnarray}
A_1 &=& \frac{A_2 + A_5}{2} = A_6\equiv A_6^{max} =
\frac{8}{9}(1 + \sqrt{v})\nonumber\\
A_{16} &=& 16-9A_6,~
A_{11}= \sqrt{A_6(16-9A_6)}.
\label{first}
\end{eqnarray}
Consistent with the inequalities (\ref{pos1}), (\ref{pos2}) and
with Eq.(\ref{state}) the remaining parameters of the covariant
copying process which do not explicitly determine the fidelity
must be chosen in the following way
\begin{eqnarray}
A_2 - A_5&=&
A_4 =   A_{3}  =  A_{7}=  A_{8}=  A_{9}  =
A_{10} =\nonumber\\&&
 A_{12}= A_{13}  = A_{14}=  A_{15} =  A_{17} = 0.
\label{rest}
\end{eqnarray}
With the help of Eq.(\ref{state}) it is straightforward to check
that for these parameters the output state $\rho_{out}(\rho_{in})$
is a non-negative operator.

Thus, consistent with the fundamental laws of quantum theory
the output state of a covariant quantum process which copies all pure two-qubit states of the same degree of entanglement $\alpha$
with the maximal fidelity $F_{max}$ is given by Eq.(\ref{output1})
with the parameters (\ref{parameters}) being determined by
Eqs.(\ref{first}) and (\ref{rest}) (compare also with Eq.(\ref{state})).

\begin{figure}
\includegraphics[width=8.cm]{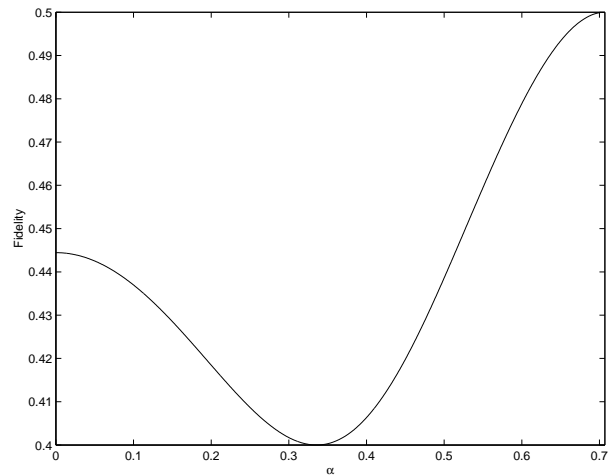}
\caption{Fidelity of the optimal covariant copying process and its
dependence on the degree of entanglement $\alpha$ of a pure
two-qubit input state. The fidelity varies between $0.4$ and
$0.5$.} \label{fidelity1}
\end{figure}

The fidelity $F_{max}$ of this optimal covariant copying process
and its dependence on the degree of entanglement $\alpha$ of
the pure two-qubit input state is depicted in
Fig.\ref{fidelity1}.
This fidelity oscillates between a minimum value of $F =0.4$ which
is assumed at $\alpha = \sqrt{1/2 - \sqrt{15}/10}$ and a maximum
value of $F= 1/2$ which is assumed at $\alpha = 1/\sqrt{2}$. The
value $\alpha =0$ corresponds to the optimal copying of two
arbitrary (generally different) separable qubit states. Consistent
with known results on optimal cloning of arbitrary single qubit
states in this latter case the fidelity $F_{max}$ assumes the
value $F= (2/3)^2$. From Fig.\ref{fidelity1} it is also apparent
for which degrees of entanglement $\alpha$ it is easier to copy
entangled states than separable ones.

At this point it is worth mentioning major differences between our
results and the previously published  results on the copying of
maximally entangled states of Ref.  \cite{Cerfetal}. In our
treatment we are interested in obtaining two separable copies, say
$|\psi \rangle \otimes |\psi\rangle$,  of entangled two-qubit
states $|\psi\rangle \in \Omega_{\alpha}$  of a given degree of
entanglement. Correspondingly, we are optimizing the two-pair
fidelity $F = \langle \psi |\otimes \langle \psi|\rho_{out}|\psi
\rangle \otimes |\psi\rangle$, for $0\leq \alpha \leq 1/\sqrt{2}$.
In contrast, in Ref.\cite{Cerfetal} the optimization of the
single-pair fidelities $F_1'=~Tr\{|\psi\rangle\langle\psi|\otimes
{\bf 1}_{34}\rho_{out}\}$ and $F_2'=~Tr\{{\bf
1}_{12}\otimes|\psi\rangle\langle\psi| \rho_{out}\}$ was
investigated for the special case of maximally entangled input
states, i.e. $\alpha = 1/\sqrt{2}$. Thus, this latter process
does not simultaneously also optimize the separability of the copies.
For maximally entangled
input states our optimized process yields for these latter figure of
merits, for example, the values $F_1'= F_2'=
0.67$ which is below the optimal value of $0.7171$ presented in Ref. \cite{Cerfetal}.
But, for the same value of $\alpha=1/\sqrt{2}$ the figure of merit of Eq.(\ref{optimalfid})
yields for our optimal copying process the values $F_{max} =0.5$
whereas the copying process of Ref. \cite{Cerfetal}
yields the smaller value of  $F_{max}=0.458$ because this latter process does not optimize with respect
to two separable copies.

\section{Optimal covariant copying processes as completely positive  quantum operations}
In this section it is demonstrated that the covariant optimal copying processes derived in Sec. IV
are completely positive deterministic quantum operations. Thus,
they  can be implemented by unitary quantum transformations with
the help of additional ancilla qubits.

Using Eqs.(\ref{output1}), (\ref{first}), (\ref{rest}) it is
straightforward to demonstrate that the output state of the
optimal covariant quantum copying process can be written in the
form
\begin{widetext}
\begin{eqnarray}
\rho_{out}(|\psi\rangle \langle \psi|
) &=& {K} |\psi\rangle \langle \psi| \otimes \frac{1}{4}{\bf 1} {K}^{\dagger} \equiv
\sum_{i,j=0,1}{\cal K}_{ij}|\psi\rangle \langle \psi| {\cal K}_{ij}
\label{out}
\end{eqnarray}
\end{widetext}
with the operators
\begin{eqnarray}
K &=&\sqrt{A_1} P^{(1,3)}_T\otimes P^{(2,4)}_T + \sqrt{A_{16}}P^{(1,3)}_S\otimes P^{(2,4)}_S
= K^{\dagger},\nonumber\\
{\cal K}_{ij} &=&\frac{K}{2} |i\rangle_3 \otimes | j\rangle_4.
\label{Kraus1}
\end{eqnarray}
Thereby, $P^{(a,b)}_T = \sum\limits_{M=0,\pm 1} |1~M\rangle
\langle 1~M | \otimes|1~M\rangle \langle 1~M | $ and $P^{(a,b)}_S
= |00\rangle \langle 00 | \otimes|00\rangle \langle 00 | $ are
projection operators onto the triplet and singlet subspaces of
qubits $a$ and $b$  and $|J~M\rangle$ denote the corresponding
(pure) two-qubit quantum states with total angular momentum
quantum numbers $J$ and magnetic quantum numbers $M$. The states
$\{|i\rangle_3; i=0,1\}$ and $\{| j\rangle_4; j=0,1\}$ denote
orthonormal basis states in the one-qubit Hilbert spaces of qubits
three and four, respectively. According to Eqs.(\ref{out}) and
(\ref{Kraus1}) the four Kraus operators \cite{Kraus} ${\cal
K}_{ij}$ ($i,j =0,1)$ characterize a quantum operation which acts
on the two input qubits one and two and which depends on their
degree of entanglement $\alpha$. These Kraus operators map
two-qubit states into four-qubit states and they fulfill the
completeness relation
\begin{eqnarray}
\sum_{i,j=0,1} {\cal K}_{ij}^{\dagger}{\cal K}_{ij} &=& {\bf 1}_{12}
\label{complete}
\end{eqnarray}
where ${\bf 1}_{12}$ denotes the unit operator in the Hilbert space of qubits one and two. Thus,
Eq.(\ref{out}) describes a deterministic quantum operation \cite{Kraus,Peres,Nielsen} acting on the two input-qubits which are to be copied
in an optimal way.
In addition, the Kraus representation of Eq.(\ref{out}) demonstrates that
the optimal covariant copying processes considered so far are completely positive quantum maps \cite{Kraus, Nielsen}.

Alternatively, the quantum operation of Eq.(\ref{out}) may also be implemented by an associated
unitary transformation ${ U}$
which involves two additional ancilla-qubits.
Denoting orthonormal basis states of these additional ancilla-qubits by $\{|\alpha\rangle_5 \otimes |\beta\rangle_6; \alpha,\beta = 0,1\}$
this unitary transformation ${ U}$ can be choosen so that it fulfills the relation
\begin{widetext}
\begin{eqnarray}
U|\psi\rangle_{12}\otimes |0\rangle_{3456} &=&\label{help1}
\sum_{i,j=0,1}\left(\frac{K}{2}|\psi\rangle_{12}\otimes
|i\rangle_{3}\otimes|j\rangle_{4}\right)\otimes|i\rangle_{5}\otimes|j\rangle_{6},
\end{eqnarray}
\end{widetext}
for example, with $|0\rangle_{3456} =|0\rangle_{3}\otimes|0\rangle_{4}\otimes|0\rangle_{5}\otimes|0\rangle_{6}$.
Thereby,
the subscripts of the state vectors label the qubits they  are referring to and the bracket of Eq.(\ref{help1})
indicates that the Kraus operators act on the system- and device-qubits only.
Due to the completeness relation (\ref{complete}) the linear transformation of Eq.(\ref{help1}) preserves scalar products, i.e.
$_{3456}\langle0|\otimes_{12}\langle\psi|U^{\dagger}U|\Phi\rangle_{12}\otimes| 0\rangle_{3456}=_{12}\langle\psi|\Phi\rangle_{12}$.
Thus, it can be completed to a unitary transformation acting on the six-qubits constituted by the system, the device-, and the
two ancilla-qubits \cite{Nielsen}.
Accordingly, the optimal covariant copying process of Eq.(\ref{out}) can be realized also with the help of
this unitary transformation $U$ in the following way: In a first step one applies the transformation $U$ to the intial state $
|\psi\rangle_{12}\otimes | 0\rangle_{3456}$
of the system-, device- and ancilla-qubits, i.e.
\begin{widetext}
\begin{eqnarray}
&&U|\psi\rangle_{12}\otimes | 0\rangle_{3456}~_{3456}\langle 0 |\otimes_{12}\langle \psi|
U^{\dagger} =
\sum_{i,j, i',j'=0,1} {\cal K}_{ij}
|\psi\rangle_{12}\otimes | i\rangle_{5}\otimes | j\rangle_{6}
~_{6}\langle j' |\otimes_{5}\langle i' | \otimes_{12}\langle \psi|
{\cal K}^{\dagger}_{i'j'}.
\end{eqnarray}
\end{widetext}
In a second step one measures
the ancilla-qubits in the orthogonal basis $\{|\alpha\rangle_5 \otimes |\beta\rangle_6; \alpha,\beta = 0,1\}$
without selection of the measurement results.
This non-selective measurement \cite{Peres} yields the output state $ \rho_{out}(|\psi\rangle \langle \psi|)$ of Eq.(\ref{out}).

\section{Properties of output states}

In this section the degree of entanglement and statistical correlations
of the output states  produced by the
optimal covariant copying processes are discussed.

As the process of copying of an arbitrary pure entangled two-qubit
state is not perfect the original and the copy will exhibit
characteristic entanglement features and statistical correlations.
Convenient measures for quantifying these properties are the
concurrences \cite{Wootters} and the indices of correlation
\cite{Barnett} of the subsystems.


Let us consider
a two-qubit state described by a density operator
$\rho$.
Its concurrence is defined in terms of the decreasing set of
eigenvalues , say
$\{\lambda_1\geq \lambda_2 \geq  \lambda_3 \geq \lambda_4\}$,
of the operator
\begin{equation}
R=\rho (\sigma _{y}\otimes \sigma _{y})\rho ^{\ast }(\sigma
_{y}\otimes \sigma _{y}).
\end{equation}
Thereby,
\begin{eqnarray}
\sigma_y &=&
\left(
\begin{array}{lr}
0 &  -i\\
i & 0
\end{array}
\right)
\end{eqnarray}
denotes the Pauli spin operator and
the star-symbol $(^*)$ denotes complex conjugation.
In terms of these eigenvalues the concurrence of the quantum state $\rho$
is defined by the relation \cite{Wootters}
\begin{equation}
\label{concurrence}
 C(\rho)= \max\Big\{0,
 \sqrt{\lambda_{1}}-\sqrt{\lambda_{2}}-\sqrt{\lambda_{3}}-\sqrt{\lambda_{4}}
 \Big\}.
\end{equation}
According to this definition the values of the concurrence are confined to
the interval $[0,1]$  with $C(\rho) = 0$ and $C(\rho) = 1$ corresponding
to a separable and a maximally entangled
two-qubit state.

A convenient measure for quantifying bipartite statistical correlations of a quantum state
$\rho$ is its index of correlation $I(\rho)$ \cite{Barnett}.
It is defined by the relation
\begin{equation}
I(\rho) =S (\rho_a) + S (\rho_b) - S (\rho);
\end{equation}
with $S(\rho_a)$, $S(\rho_b)$, and $S(\rho)$ denoting the von
Neumann-entropies of subsystems $a$, $b$ and of the whole system.
The corresponding reduced density operators of the subsystems are
denoted by $\rho_j = {\rm Tr}_{i\ne j} (\rho)$ with $j =a,b$.
Correspondingly,
the index of correlation $I (\rho)$ vanishes for all
uncorrelated (factorizable) states and it attains its largest value for maximally
entangled pure states (with $\alpha =1/\sqrt{2}$).
Let us now investigate
entanglement
and statistical correlations
of the output state $\rho_{out}(\rho_{in})$ with respect to
the first and the second qubit,
with respect to the first and third qubit
and with respect to qubits one and two on the one hand and qubits three and four on the other hand.

\subsection{Entanglement and statistical correlations of qubits one and two}

Let us consider first of all a two-qubit input state of the form $\rho_{in} = |\psi\rangle \langle \psi |$
with
$|\psi\rangle \in \Omega_{\alpha}$.
Both its concurrence as well as its index of correlation are given by
\begin{eqnarray}
C(\rho_{in})&=& 2 |\alpha\sqrt{1-\alpha^2}|,\\
I(\rho_{in})&=& -2 \left\{ |\alpha|^2\ln{|\alpha|^2} +
|\beta|^2\ln{|\beta|^2} \right\}
\end{eqnarray}
with $\beta = \sqrt{1-\alpha^2}$.
The corresponding reduced density operator $\rho_{out}^{(1,2)}$ of qubits one and two after
an optimal covariant copying process can be determined easily from Eqs.(\ref{output1}),
(\ref{parameters}),(\ref{first}), and (\ref{rest}). In particular, its concurrence, for example, is given by
\begin{equation}
C(\rho_{out}^{(1,2)}) = \frac{1}{16} \max \left\{0,
(4|\alpha\beta|+1)(2A_6+A_{11}) - 8\right\}.
\end{equation}
The corresponding index of correlation can also be evaluated easily.
Due to the inherent symmetry of the optimal covariant copying process the reduced density operators
of qubits one and two on the one hand and qubits three and four on the other hand are equal. Therefore,
all results obtained for the system-qubits one and two are also valid for
the device-qubits three and four.

\begin{figure}
\includegraphics[width=8.cm]{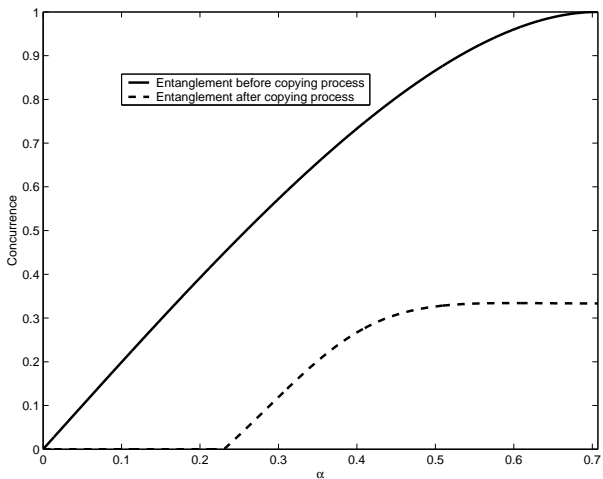}
\caption{The dependence of the concurrence of the quantum states
of qubits one and two before (solid line) and after (dashed line)
the optimal covariant copying process on the degree of
entanglement ${\alpha}$. $\alpha =0$ and $\alpha = 1/\sqrt{2}$
correspond to the limits of a separable and a maximally entangled
pure two-qubit input state.} \label{conc1}
\end{figure}

In Fig.\ref{conc1} the concurrence of the quantum states of qubits
one and two before and after the optimal covariant copying process
are depicted.
The concurrence
of the pure input state increases smoothly from its minimum value zero at $\alpha =0$
to its maximum value of unity at $\alpha =1/\sqrt{2}$.
The corresponding values of the output states
with respect to qubits one and two exhibit a rather different
behavior.
First of all, it is apparent that a minimum degree of entanglement
$\alpha_{min} =0.231$ of the pure input state $\rho_{in}$ is
required in order to achieve also an entanglement between qubits
one and two in the resulting output state. Secondly, the
concurrence of the output state saturates at a rather moderate
value around $0.3$ at which it becomes almost independent of the
value of $\alpha$. Thirdly, the maximum entanglement between
qubits one and two is not achieved exactly for maximally entangled
initial states with $\alpha =1/\sqrt{2}$ but for values slightly
below. However, this difference is very small.

\begin{figure}
\includegraphics[width=8.cm]{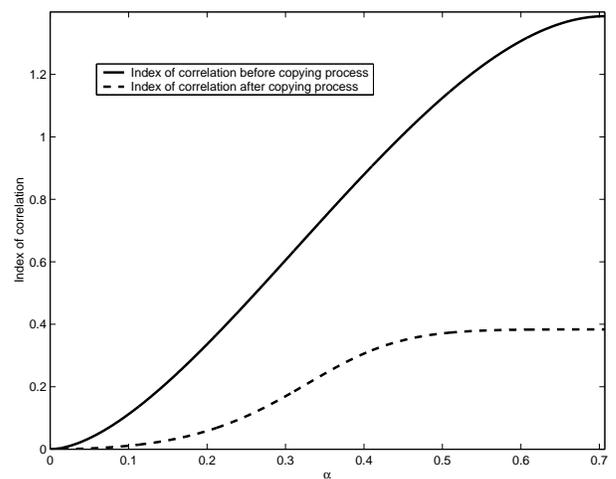}
\caption{Indices of correlations of input and output states with
respect to qubits one and two and their dependence on the degree
of entanglement $\alpha$ of the pure input state. As in the case
of concurrence the correlations saturate at a rather moderate
level.} \label{index1}
\end{figure}

The corresponding indices of correlations and their dependence on the degree of entanglement of
input and output states are depicted in Fig.\ref{index1}.
In contrast to the concurrence the index of correlation of the output state increases smoothly
with increasing values of $\alpha$ and reaches its maximum exactly at $\alpha=1/\sqrt{2}$.
Similar to the case of the concurrence
there is a considerable drop of the index
of correlation of the copied pair in comparison with its original.

\subsection{Correlation of the first and third qubit}

\begin{figure}
\includegraphics[width=8.cm]{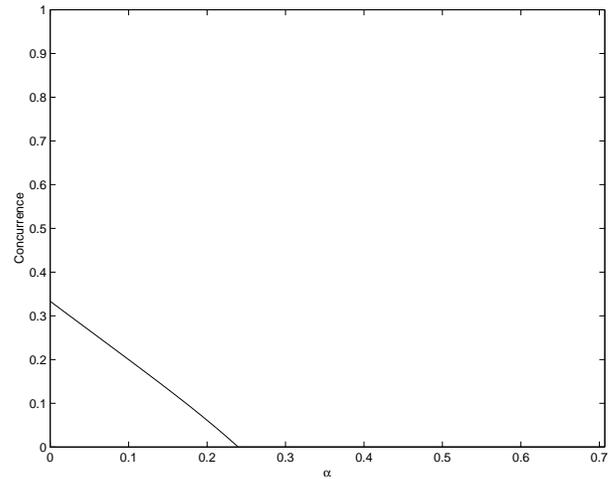}
\caption{Concurrence of the output state of an optimal covariant
copying process with respect to qubits one and three and its
dependence on the degree of entanglement $\alpha$ of the pure
two-qubit input state. For $\alpha \ge 0.241$ the concurrence
vanishes.} \label{Conc2}
\end{figure}

\begin{figure}
\includegraphics[width=8.cm]{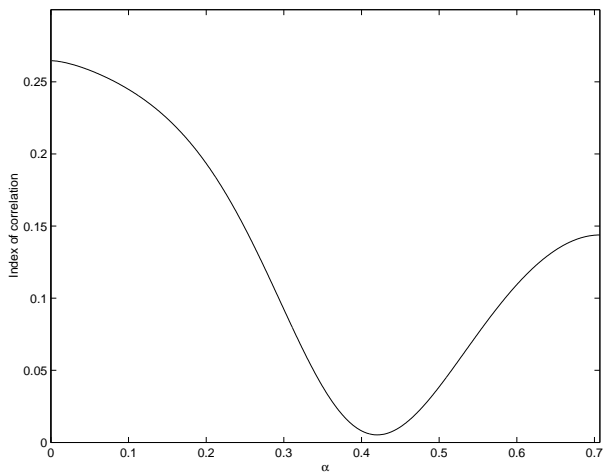}
\caption{Index of correlation of the output state of an optimal covariant copying process with respect to qubits one and three
and its dependence on the degree of entanglement
$\alpha$ of the pure two-qubit input state.}
\label{index2}
\end{figure}

In view of the structure of the input state $\rho_0$ of Eq.(\ref{inputdm}) the entanglement and statistical
correlation between qubits one and three vanish.
The concurrence of the
reduced
density operator of the output state of the optimal covariant copying process  with respect to  these qubits
is given by
\begin{equation}
C_{13}^{(out)}= \frac{1}{4}\max\{0, |-4+3A_6|-3A_6\alpha
\sqrt{1-\alpha^2}\}.
\end{equation}
This concurrence and the corresponding index of correlation of the
output state and their dependence on the degree of entanglement
$\alpha$ of the input state are depicted in Figs.\ref{Conc2} and
\ref{index2}.
Characteristically, the concurrence decreases
linearly from its maximum value at $\alpha =0$
until it vanishes for
$\alpha \geq  0.241$.
Contrary to the concurrence the index of correlation depends smoothly on the degree of entanglement $\alpha$
of the pure input state.
Furthermore, it decreases up to the
value $\alpha \approx 0.421$ where it assumes a minimum. For degrees of entanglement $\alpha \geq 0.421$ it increases
monotonically and reaches a local maximum at $\alpha = 1/\sqrt{2}$ which corresponds to a maximally entangled pure input state.

\subsection{Correlation between two copies}

\begin{figure}
\includegraphics[width=8.cm]{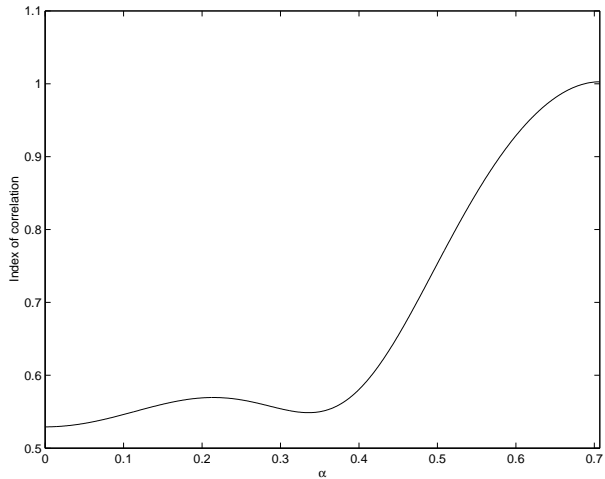}
\caption{Index of correlation of the output state of an optimal
covariant copying process with respect to qubits one and two  and
qubits three and four and its dependence on the degree of
entanglement $\alpha$ of the pure two-qubit input state.}
\label{index3}
\end{figure}

Finally, let us discuss the statistical correlations of the output state
with respect to qubits one and two on the one hand
and qubits three and four on the other hand.
Its dependence on the degree of entanglement $\alpha$ of the pure two-qubit input state
is depicted in Fig.\ref{index3}.
\begin{figure}
\includegraphics[width=8.cm]{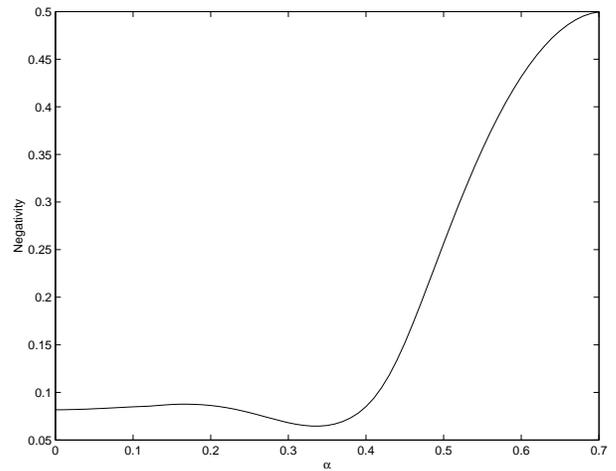}
\caption{Negativity of the output state of an optimal covariant
copying process and its dependence on the degree of entanglement
$\alpha$ of the pure two-qubit input state. The negativity is for
all values of $\alpha$ positive a hence the outputs states are
always entangled.} \label{negativity}
\end{figure}
Characteristically, one notices two maxima and two minima. The
local minimum at $\alpha = \sqrt{1/2 - \sqrt{15}/10}$ corresponds
to the optimal covariant copying process with the smallest
fidelity of the output copies (compare with  Fig.\ref{fidelity1}).
The global maximum corresponds to the copying of maximally
entangled pure input states with $\alpha = 1/\sqrt{2}$.

To conclude our discussion of entanglement and correlations
let us consider the negativity of the output state $\rho_{out}(\rho_{in})$. This quantity allows
to decide whether a quantum state contains free entanglement or not.
The definition of the negativity of a quantum state starts from the observation, that
the partial transpose of a separable state always yields a positive density operator.
The negativity is defined by the sum
of absolute values of the negative eigenvalues of the partially
transposed density operator \cite{Vidal}. In Fig.\ref{negativity}
the negativity of the output state is depicted. Thereby, the four-qubit output state is considered as
a bipartite state with respect to the system-qubits one and two and the device-qubits three and four.
The dependence of this negativity
on the degree of entanglement $\alpha$ of the input states resembles the corresponding
dependence of the correlation function
presented in Fig.\ref{index3}. Fig.\ref{negativity} illustrates several interesting
features: Firstly, the output state is
entangled for all values of $\alpha$. Secondly,
the global minimum of the negativity coincides with the point of
worst copying, i.e. with $\alpha = \sqrt{1/2 - \sqrt{15}/10}$. The maximum
at the point $\alpha=1/\sqrt{2}$ indicates that the copying of
maximally entangled states results in a maximally entangled output
state.

\section{Conclusion}
We investigated the copying of pure entangled two-qubit states of a given degree of entanglement.
Optimizing these processes with respect to two separable copies
we identified
the optimal covariant and completely positive copying processes for all possible degrees of entanglement.
It was demonstrated that
the fidelity of the resulting output states with respect to separable copies varies between
the values of $0.4$  and $0.5$. In particular, this latter value characterizes the optimal
copying of maximally entangled two-qubit states.
In the special case of factorizable input states  we
obtain the already known value of $4/9$. We investigated
correlation properties and the entanglement of the resulting output states.
We want to point out that the
presented approach which is based on an analysis of the irreducible tensor components of the input state
may be generalized also to more complex situations, such as
the copying of $N$ entangled pairs to $M$ pairs.  Work along these lines is in progress.

\acknowledgements

Financial support by GA\v CR 202/04/2101, VZ M\v SMT 210000018 of
the Czech Republic, by the Alexander von Humboldt foundation and
by the Deutsche Forschungsgemeinschaft (SPP QUIV) is gratefully
acknowledged.

\newpage
\appendix

\section{Optimal copying processes and covariant quantum maps}

The proof that any optimal quantum copying process $\widehat{T_{\alpha}}$
of the form (\ref{map})
can be represented by a corresponding covariant quantum map of the form of Eq.(\ref{covariance})
with
the same fidelity is similar to the proof given by Werner in the context of optimal
cloning of arbitrary $d$-dimensional quantum states \cite{Werner}.
The only major difference concerns the group
operations which connect all possible pure input states. In our case this group involves all unitary transformations of the form
$U_1\otimes U_2$ with $U_j \in SU_2$.
For the sake of completeness we briefly outline the main steps involved in this proof.

Let us start from
the definition of the optimal fidelity of our copying process which is given by
\begin{equation}
\widehat{{\mathcal{L}}_{\alpha}} = \sup_{T_{\alpha}}
{\mathcal{L}}(T_{\alpha})= \sup_{T_{\alpha}}\left\{\inf_{|\psi \rangle
\in \Omega_{\alpha}} \underbrace{\langle \psi | \otimes \langle \psi |
\rho_{out} | \psi \rangle \otimes | \psi \rangle}_{Z_{\psi}} \right\}.
\end{equation}
The functions ${Z_{\psi}}$ are continuous. Largest lower bounds (infima) of
a set of continuous functions yield upper semi-continuous functions.
In a finite dimensional Hilbert space
the set of admissible quantum operations $T_{\alpha}$ is closed, and bounded. Therefore
it constitutes a compact set. However, an upper semicontinuous function on a compact set must
acquire its maximum. Let us denote this maximum
$\widehat{T_{\alpha}}$. Thus
\begin{equation}
\widehat{{\mathcal{L}}_{\alpha}} = \sup_{T_{\alpha}}
{\mathcal{L}}(T_{\alpha}) = {\mathcal{L}}(\widehat{T}_{\alpha}).
\end{equation}

For an arbitrary admissible quantum copying map $T_{\alpha}$ its average $\tilde{T}_{\alpha}$ over all group operations
 is defined by
\begin{equation}
\tilde{T}_{\alpha} (\rho) = \int dU_{1} dU_{2}
{\cal U}
^{*\otimes 2} T_{\alpha}\left({\cal U}
 \rho {\cal U}^* \right) {\cal U}^{\otimes 2}
\end{equation}
with ${\cal U} = U_1 \otimes U_2$.
Thereby,
$dU_{1}dU_{2}$ denotes the normalized left invariant Haar measure of the group
$SU(2)\times SU(2)$. The map $\tilde{T}_{\alpha}$ is again an
admissible copying map, which fulfills the covariance condition
(\ref{covariant}).
If $\tilde{T}_{\alpha}$ denotes the average of the
optimal copying process $\widehat{T}_{\alpha}$,  then for every pure state
$|\psi \rangle \in \Omega_{\alpha}$ we find
\begin{widetext}
\begin{eqnarray}
&&
\langle \psi | \otimes \langle \psi |
\tilde{T}_{\alpha}(|\psi \rangle \langle \psi|) | \psi \rangle \otimes | \psi \rangle =
\int dU_{1}dU_{2}\langle \psi_{{\cal U}} | \otimes \langle \psi_{{\cal U}} |
\tilde{T}_{\alpha}\left( |\psi_{{\cal U}}
 \rangle \langle \psi_{{\cal U}}| \right) | \psi_{{\cal U}} \rangle \otimes | \psi_{{\cal U}} \rangle
\geq \int dU_{1}dU_{2}{\mathcal{L}}(\widehat{T}_{\alpha}) = \widehat{{\mathcal{L}}}_{\alpha}
\end{eqnarray}
\end{widetext}
with $|\psi_{{\cal U}} \rangle = {\cal U} |\psi \rangle$.
Because the lefthand side of the inequality is
independent of $|\psi \rangle \in
\Omega_{\alpha}$, it is also valid for
${\mathcal{L}}(\tilde{T}_{\alpha})$, i.e.
\begin{equation}
{\mathcal{L}}(\tilde{T}_{\alpha}) \geq \widehat{{\mathcal{L}}}_{\alpha}.
\end{equation}
But
from the definition of $ \widehat{{\mathcal{L}}}_{\alpha}$ we know
that ${\mathcal{L}}(\tilde{T}_{\alpha}) \leq
\widehat{{\mathcal{L}}}_{\alpha}$. Thus, we conclude that
\begin{equation}
{\mathcal{L}}(\tilde{T}_{\alpha}) = \widehat{{\mathcal{L}}}_{\alpha}
\end{equation}
and
we can restrict our search for optimal copying processes to
covariant ones.

\section{Irreducible tensor operators}

A set of irreducible tensor operators $T^{(a,b)}(J_{1}J_{2})_{KQ}$ for two quantum systems
$a$ and $b$ with respect to the rotation group is defined by \cite{Blum, Biedenharn}
\begin{widetext}
\begin{eqnarray}
\label{irreducible}
T^{(a,b)}(J_{a},J_{b})_{K Q} &=&\sum_{M_{a} M_{b}}(-1)^{J_{a}-M_{a}}\sqrt{2K+1}
\left(
\begin{array}{lcr}
J_{a} & J_{b} & K\\
M_{a} & -M_{b} & -Q
\end{array}
\right)
|J_{a} M_{a}\rangle \otimes \langle J_{b} M_{b}|
\end{eqnarray}
\end{widetext}
with the ket (bra) $|J M \rangle$ ($\langle J M |$)
denoting eigenstates of the total angular momentum operator of both quantum systems.
Thereby the total angular momentum quantum number and the magnetic quantum number are
denoted by $J$ and $M$. In Eq.(\ref{irreducible}) we have introduced the
3j-symbol whose orthogonality and completeness relations
imply the corresponding relations
\begin{widetext}
\begin{eqnarray}
{\rm Tr}[T^{(a,b)}(J_a,J_b)_{K Q}T^{(a,b)}(J_a',J_b')^{\dagger}_{K' Q'}]&=&
\delta_{J_a J_a'} \delta_{J_b J_b'} \delta_{K K'} \delta_{Q Q'}.
\end{eqnarray}
\end{widetext}
Thereby,
${\rm Tr}$ denotes the trace over the Hilbert spaces of quantum systems
$a$ and $b$.
The
irreducible tensor operators of Eq.(\ref{irreducible}) are special cases of complete
orthogonal sets of operators with simple transformation
properties with respect to a particular group. In the case of
Eq.(\ref{irreducible}) it is the group of rotations of the quantum systems
$a$ and $b$ and the simple transformation property is given by
\begin{widetext}
\begin{eqnarray}
{\cal U}
T^{(a,b)}(J_a J_b)_{KQ}
{\cal U}^{\dagger}
&=&
\sum_{q}
T^{(a,b)}(J_a J_b)_{Kq}
D(U)_{qQ}^{(K)}.
\label{irred2}
\end{eqnarray}
\end{widetext}
with ${\cal U} = U\otimes U$.

As the tensor operators of Eq.(\ref{irreducible}) form a complete set
any operator $\rho$ in the Hilbert space of particles $a$ and $b$ can be decomposed according to
\begin{equation}
\rho= \sum_{J_a J_b K Q} \left\langle T^{(a,b)}(J_a J_b)^{\dagger}_{KQ} \right\rangle
T^{(a,b)}(J_a J_b)_{KQ}
\label{decompose}
\end{equation}
with
\begin{eqnarray}
\label{help}
\left\langle T^{(a,b)}(J_a J_b)^{\dagger}_{KQ} \right\rangle &=& Tr \left\{
\rho~T^{(a,b)}(J_a J_b)^{\dagger}_{KQ} \right\},\\
T^{(a,b)}(J_a J_b)^{\dagger}_{K Q}
&=& (-1)^{J_a - J_b + Q}~
 T^{(a,b)}(J_b J_a)_{K -Q}.\nonumber
\end{eqnarray}
With the help of these relations and
the condition $ U_1\otimes U_2\rho_{ref} U_1^{\dagger}\otimes U_2^{\dagger} = \rho_{ref}$
which has to be fulfilled by any covariant quantum process
it is straightforward to proof the general form of the output state of  Eq.(\ref{output1}).

\section{Positivity constraints}

We start from the most general form of the output state of the linear and covariant quantum map defined by
Eqs.(\ref{output1}) and (\ref{parameters}). Due to the covariance condition (\ref{covariance}) this output
state can be decomposed into a direct sum of density operators according to
\begin{eqnarray}
\rho_{out}(\rho_{in}) &=& M_1 \oplus M_2 \oplus M_3 \oplus M_4 \oplus M_5
\label{matrix}
\end{eqnarray}
with
\begin{widetext}
\begin{eqnarray}
M_1 &=& [(2\alpha^2 -1)(A_2 + A_5) + A_1 + A_6]|1 1; 1 1\rangle
\langle 1 1; 1 1| + A_6 |1 0; 1 0\rangle \langle 1 0; 1 0| + A_8
|1 0; 0 0\rangle \langle 1 0; 0 0| + A_{16} |0 0; 0 0\rangle
\langle 0 0; 0 0| +\nonumber\\&& A_{14} |0 0; 1 0\rangle \langle 0
0; 1 0| + [(1-2\alpha^2)(A_2 + A_5) + A_1 + A_6]|0 0; 1 -1\rangle
\langle 0 0; 1 -1| + 2\alpha \sqrt{1-\alpha^2}[A_1 |1 1; 1
1\rangle \langle 1 0; 1 0| +\nonumber\\&& A_1|1 0; 1 0\rangle
\langle 1 1; 1 1|] - 2\alpha \sqrt{1-\alpha^2}[A_3|1 1; 1 1\rangle
\langle 1 0; 0 0| + A_3^*|1 0; 0 0\rangle \langle 1 1; 1 1|] +
\nonumber\\ && 2\alpha \sqrt{1-\alpha^2}[A_{11}[|1 1; 1 1\rangle
\langle 0 0; 0 0| + A_{11}^*|0 0; 0 0\rangle \langle 1 1; 1 1|] -
2\alpha \sqrt{1-\alpha^2}[A_{9}[|1 1; 1 1\rangle \langle 0 0; 1 0|
+ A_9^*|0 0; 1 0\rangle \langle 1 1; 1 1|] + \nonumber\\ &&
(2\alpha^2 -1)[A_7 |1 0; 1 0\rangle \langle 1 0; 0 0| + A_7^*|1
0;0 0\rangle \langle 1 0; 1 0|] +
[A_{11}|1 0; 1 0\rangle \langle 0 0; 0 0| + A_{11}^*|0 0;0 0\rangle \langle 1 0; 1 0|] +\nonumber\\
&& (2\alpha^2 -1)[A_{10} |1 0; 1 0\rangle \langle 0 0; 1 0| +
A_{10}^*|0 0;1 0\rangle \langle 1 0; 1 0|] + 2\alpha
\sqrt{1-\alpha^2}[A_{1}|1 0; 1 0\rangle \langle 0 0; 1 -1| +
A_{1}|0 0;1 -1 \rangle \langle 1 0; 1 0|] +\nonumber\\&&
(2\alpha^2 -1)[A_{12} |1 0; 0 0\rangle \langle 0 0; 0 0| + A_{12}
^*|0 0;0 0\rangle \langle 1 0; 0 0|] + [A_{17}|1 0; 0 0\rangle
\langle 0 0; 1 0| + A_{17}^*|0 0;1 0\rangle \langle 1 0; 0 0|] +
\nonumber\\ && 2\alpha \sqrt{1-\alpha^2}[A_{3}^*|1 0; 0 0\rangle
\langle 0 0; 1 -1| + A_{3}|0 0;1 -1 \rangle \langle 1 0; 0 0|] +
(2\alpha^2 -1)[A_{15}^* |0 0; 0 0\rangle \langle 0 0; 1 0| +
A_{15}|0 0;1 0\rangle \langle 0 0; 0 0|] + \nonumber\\ && 2\alpha
\sqrt{1-\alpha^2}[A_{11}^*|0 0; 0 0\rangle \langle 0 0; 1 -1| +
A_{11}|0 0;1 -1\rangle \langle 0 0; 0 0|] +\nonumber\\
&& 2\alpha
\sqrt{1-\alpha^2}[A_{9}^*|0 0; 1 0\rangle \langle 0 0; 1 -1| +
A_{9}|0 0;1 -1\rangle \langle 0 0; 1 0|],
\nonumber\\
M_2 &=& [(2\alpha^2 -1)A_4 + A_8)]|1 1; 0 0\rangle \langle 1 1; 0
0| + [-(2\alpha^2 -1)A_5 + A_6)]|1 0; 1 -1\rangle \langle 1 0; 1
-1| + \nonumber\\ && [-(2\alpha^2 -1)A_{13} + A_{14})]|0 0; 1
-1\rangle \langle 0 0; 1 -1| + [(2\alpha^2 -1)A_{2} + A_{6})]|1 1;
1 0\rangle \langle 1 1; 1 0| + \nonumber\\ && 2\alpha
\sqrt{1-\alpha^2}[A_{3}^*|1 1; 0 0\rangle \langle 1 0; 1 -1| +
A_{3}|1 0;1 -1\rangle \langle 1 1; 0 0|] -
\nonumber\\&&
2\alpha
\sqrt{1-\alpha^2}[A_{17}|1 1; 0 0\rangle \langle 0 0; 1 -1| +
A_{17}^*|0 0;1 -1\rangle \langle 1 1; 0 0|] + 
[(2\alpha^2 -1)A_{7}^* + A_3^*)]|1 1; 0 0\rangle \langle 1 1; 1 0|
+\nonumber\\ &&
 [(2\alpha^2 -1)A_{7} + A_3)]|1 1; 1 0\rangle \langle 1 1; 0 0| +
[(2\alpha^2 -1)A_{10} - A_9]|1 0; 1 -1\rangle \langle 0 0; 1 -1| +\nonumber\\
&& [(2\alpha^2 -1)A_{10}^* - A_9^*]|0 0; 1 -1\rangle \langle 1 0;
1 -1| + 2\alpha \sqrt{1-\alpha^2}[A_{1}|1 0; 1 -1\rangle \langle 1
1; 1 0| + A_{1}|1 1;1 0\rangle \langle 1 0; 1 -1|] - \nonumber\\
&& 2\alpha \sqrt{1-\alpha^2}[A_{9}^*|0 0; 1 -1\rangle \langle 1 1;
1 0| + A_{9}|1 1;1 0\rangle \langle 0 0; 1 -1|],
\nonumber\\
M_3 &=& [-(2\alpha^2 -1)A_4 + A_8)]|1 -1; 0 0\rangle \langle 1 -1;
0 0| + [(2\alpha^2 -1)A_5 + A_6)]|1 0; 1 1\rangle \langle 1 0; 1
1| + \nonumber\\ && [(2\alpha^2 -1)A_{13} + A_{14})]|0 0; 1
1\rangle \langle 0 0; 1 1| + [-(2\alpha^2 -1)A_{2} + A_{6})]|1 -1;
1 0\rangle \langle 1 -1; 1 0| - \nonumber\\ && 2\alpha
\sqrt{1-\alpha^2}[A_{3}^*|1 -1; 0 0\rangle \langle 1 0; 1 1| +
A_{3}|1 0;1 1\rangle \langle 1 -1; 0 0|] -\nonumber\\&&
2\alpha
\sqrt{1-\alpha^2}[A_{17}|1 -1; 0 0\rangle \langle 0 0; 1 1| +
A_{17}^*|0 0;1 1\rangle \langle 1 -1; 0 0|] + 
[(2\alpha^2 -1)A_{7}^* - A_3^*]|1 -1; 0 0\rangle \langle 1 -1; 1
0| +\nonumber\\ &&
  [(2\alpha^2 -1)A_{7} - A_3]|1 -1; 1 0\rangle \langle 1 -1; 0
0| +  \nonumber \\ && [(2\alpha^2 -1)A_{10} + A_9]|1 0; 1 1\rangle
\langle 0 0; 1 1| +
 [(2\alpha^2 -1)A_{10}^* + A_9^*]|0 0; 1 1\rangle \langle 1 0; 1 1|
 + \nonumber \\ &&
2\alpha \sqrt{1-\alpha^2}[A_{1}|1 0; 1 1\rangle \langle 1 -1; 1 0|
+A_{1}|1 -1;1 0\rangle \langle 1 0; 1 1|] +  \nonumber \\
&& 2\alpha \sqrt{1-\alpha^2}[A_{9}^*|0 0; 1 1\rangle \langle 1 -1;
1 0| + A_{9}|1 -1;1 0\rangle \langle 0 0; 1 1|], \nonumber
\\
M_4 &=&
[(2\alpha^2 -1)(A_2 - A_5) - A_1 + A_6]|1 1; 1 -1\rangle \langle 1 1; 1 -1|,\nonumber\\
M_5 &=& [(2\alpha^2 -1)(-A_2 + A_5) - A_1 + A_6]|1 -1; 1 1\rangle
\langle 1 -1; 1 1|. 
\label{state}
\end{eqnarray}
\end{widetext}
Thereby,
the basis states $|JM; J'M'\rangle$ involve eigenstates of the total angular momenta of qubits one and three on the one hand
and qubits two and four on the other hand, i.e.
$|JM; J'M'\rangle = |JM\rangle_{(1,3)} \otimes |J'M'\rangle_{(2,4)}$ with $(J,M)$ and $(J',M ')$ denoting the relevant total angular momentum
and magnetic quantum numbers.

The non-negativity of the output state (\ref{state}) necessarily implies that all diagonal matrix elements
have to be non-negative. The resulting constraints give rise to the inequalities (\ref{pos1}).
Furthermore,
for appropriately chosen pure states $| \chi \rangle$
the relation $\langle \chi | \rho_{out}(\rho_{in}) | \chi \rangle \geq 0$
yields the inequalities (\ref{pos2}), i.e.
\begin{eqnarray}
| \chi \rangle
&=& a |1 0; 1 0 \rangle + b |0 0; 0 0 \rangle ~\to~\mid A_{11}\mid^2 \leq A_{16}A_6,\nonumber\\
| \chi \rangle
&=& a |1 0; 0 0 \rangle + b |0 0; 1 0 \rangle ~\to~\mid A_{17}\mid^2 \leq A_{14}A_8,\nonumber\\
| \chi \rangle
&=& a |1 1; 1 1 \rangle + b |0 0; 1 -1 \rangle + c|1 0; 1 0 \rangle\nonumber\\
&\to&
A_6 \mid (2\alpha^2 -1)(A_2 + A_5)\mid^2 \leq
(A_1 + A_6)^2A_6 -\nonumber\\
&& 8\alpha^2 (1-\alpha^2) A_1^2 (A_1 + A_6)
\label{conditions}
\end{eqnarray}
with $a$, $b$ and $c$ denoting arbitrary complex-valued coefficients.

\end{document}